\documentclass[12pt]{article}

\usepackage{amsmath}
\usepackage{euscript}

\usepackage{pstricks,epsfig,amssymb,axodraw}
\usepackage{cite}  

\def\CCol{{\tt SANC}}

\begin{document}

\begin{titlepage}
\thispagestyle{empty}

\begin{flushright}
{\bf  
CERN-TH/2003-075\\
} 
\end{flushright}
 
\noindent
\vspace*{5mm}
\begin{center}
{
     \Large\bf How to use {\tt SANC} to improve \\
     \Large\bf the {\tt PHOTOS} Monte Carlo simulation  \\ 
     \Large\bf of bremsstrahlung in   leptonic $W$-Boson decays  }
\end{center}

\vspace*{2mm} 
\begin{center}
  {\bf G. Nanava$^{a}$ and}~~
  {\bf Z. W\c{a}s$^{b,c}$ }
  \\
  \vspace{5mm}
  {\em $^a$ Lab. of Nuclear Problems, JINR, RU-141980 Dubna, Russia}\\
  \vspace{1mm}
  {\em $^b$ Institute of Nuclear Physics,
           ul. Radzikowskiego 152, PL-31-342 Cracow, Poland}\\
  \vspace{1mm}
  {\em $^c$  CERN, Theory Division, CH-1211 Geneva 23, Switzerland }
\end{center}
\vfill

\vspace*{5mm} 
\begin{abstract}
Using the {\tt SANC} system we study the one-loop electroweak standard model
predictions, including virtual and real photon emission, for the decays
of the on-shell vector boson, $W \to l {\nu_l} (\gamma)$. 
The complete one-loop corrections and exact photon emission
matrix element are taken into account. For the 
phase-space integration, the  Monte Carlo technique is used.
This provides a useful element, first  for the evaluation of the theoretical 
uncertainty of {\tt PHOTOS}. Later we analyse the source of the differences
between  {\tt SANC} and {\tt PHOTOS} and we calculate the additional weight,
which once installed, improves predictions of {\tt PHOTOS} simulations.
 We can conclude that, after the correction of the  weight is
implemented, the  theoretical uncertainty  of  {\tt PHOTOS} simulations 
due to an incomplete first-order matrix element is reduced to  below 
$ { \alpha \over \pi} $, for observables not
 tagging the photon in a direct way, and to 10\% otherwise. This is interesting for 
applications in  the phenomenology of the ongoing LEP2 and future
 LC and LHC experimental studies.
\end{abstract}  
\vspace*{1mm}

\begin{center}
  {\em Submitted to Acta Physica Polonica}
\end{center}

\vspace*{1mm}
\vfill
\begin{flushleft}
{\bf CERN-TH/2003-075 \\
  March 2003 }
\end{flushleft}

\vspace*{1mm}
\bigskip
\footnoterule
\noindent
{\footnotesize \noindent
Work supported in part by the European Union 5-th Framework under contract HPRN-CT-2000-00149,
Polish State Committee for Scientific Research 
(KBN) grant 2 P03B 001 22, 
NATO grant PST.CLG.977751
and INTAS N$^{o}$ 00-00313. This paper includes material presented at the  
Cracow Epiphany Conference, January 2003.
}
\end{titlepage}

\normalsize
\newpage

\section{Introduction}

The \CCol\ project of Ref.~\cite{Bardin:2002am} has several purposes.
The intermediate  goal is to summarize
and consolidate the effort of the last three decades  in calculating
Standard Model radiative corrections for LEP, in a well organized 
 calculational environment, for future reference.
However, it is aimed not only at training young 
researchers and students, but at some remaining
calculational projects for LEP as well. Recently it was used for 
that purpose in ~\cite{Gizo}.

An important lesson from the LEP experiments \cite{Kobel:2000aw}
is that the desirable
way of providing theoretical predictions is in the form of  Monte Carlo
event generators.
This aspect has been taken into account in the development of \CCol\
from an early stage of its development.

The currently available  version   can construct
one-loop spin amplitudes for the
decays of the gauge bosons $W$ and $Z$ and of the Higgs boson $H$.
For the moment, \CCol\ features single real-photon emission, in
the calculations of the total rate and
decay spectra of the $B \to f {\bar f} (\gamma)$ process.
The complete spin polarization density matrix of the decaying boson
is  taken into account as well.

The integration, with the Monte Carlo method, 
over the three- (two)-body final state is done without 
any approximation (in particular, the  small-mass approximation is not used).
The program provides MC events with constant weight (unweighted events).
The whole system is, therefore, fairly self-contained and complete.%

It is of the utmost importance for such a system  to
reproduce known  results precisely; this step of its development was completed in 
ref.~\cite{Gizo}. Let us turn now to its application:
first find an approximation of the relatively large terms missing in  
{\tt PHOTOS} \cite{Barberio:1994qi} for generation of  bremsstrahlung 
in $W$ decay, later install them into this program,
 and finally verify, again with \CCol{},    that
the {\tt PHOTOS} algorithm for the generation of bremmstrahlung  in decays,
 after modifications, works 
indeed better. 

Our paper is organized as follows: in the next section  we recall,
from \cite{Gizo}, 
the set of observables chosen for tests and we explain 
the input parameters used in  \CCol\ and
 {\tt PHOTOS}.
Section 3 is devoted to a discussion of  the \CCol\ matrix elements for 
 QED bremsstrahlung. 
The gauge-invariant part of this matrix element, which is expected
to be responsible for the  {\tt PHOTOS} discrepancies with  exact 
matrix element, is removed from   \CCol{} first. 
These  terms are approximated by  simple formula and
installed to a {\tt PHOTOS}.
Section 4 describes  comparisons
of  full \CCol{} with {\tt PHOTOS} with the new correcting
weight introduced.
A summary, section 5, closes the paper.

\section{Initialization set-ups for \CCol\ and {\tt PHOTOS} runs}

In the following sections we compare predictions from the programs
\CCol \, and {\tt PHOTOS}. It is essential that the initialization  
be identical in all cases and close to the physical reality;
in particular the following options are set in the two programs:
\begin{itemize}
\item
  In \CCol\ we switch off the EW part of the radiative corrections.
  The soft/hard photon limit is kept at 0.005 of the decaying particle mass. 
\item
  In {\tt PHOTOS} we switch off the double bremsstrahlung corrections.
  The soft/hard photon limit is kept at
  0.005 of the decaying particle mass.
  For the generation of the Born-level two-body decays,
   we use the Monte Carlo generation from
  \CCol{}.
\end{itemize}

To visualize the differences (or the agreement) between the calculations,
we choose a certain class of (pseudo-)observables, more precisely 
the one-dimensional distributions, which  are quite similar to the ones
used in the  first tests of {\tt PHOTOS} reported in \cite{Barberio:1991ms}. 
To visualize the usually small differences,
we plot ratios of the predictions from the two programs rather 
than the distributions themselves.

\vskip 2 mm

List of observables:
\begin{itemize}
\item
{\bf  -A-} {\it Photon energy in the decaying particle rest frame:} this
observable is sensitive mainly to the leading-log (i.e. collinear) non-infrared
(i.e. not soft) component of the distributions.
\item
{\bf  -B-}  {\it Energy of the final-state charged particle:} as  the 
previous one, this
observable is sensitive mainly to the leading-log (i.e. collinear) non-infrared
(i.e. not soft) component of the distributions.
\item
{\bf  -C-}  {\it Angle of the photon with  
final-state charged particle:} this
observable is sensitive mainly to the non-collinear  (i.e. non-leading-log) 
but soft 
(i.e. infrared) component of the distributions.
\item
{\bf  -D-}  {\it Acollinearity angle of the final-state charged particles:} this
observable is sensitive mainly to the non-collinear  (i.e. non-leading-log) 
and  non-soft 
(i.e. non-infrared) component of the distributions.
\end{itemize}

\section{Complete and truncated matrix elements of    \CCol\ }

 The Feynman diagrams for the  decay 
\begin{eqnarray}
  W^{\pm}(Q,\lambda) \rightarrow l(p_{l},\lambda_{l}) + \bar{\nu}(p_{\nu},\lambda_{\nu}) + \gamma(k,\sigma)
\end{eqnarray}
 are shown in fig.~1 (unitary gauge), where $(p_{l},\lambda_{l})$ and $(p_{\nu},\lambda_{\nu})$ denote
 the four-momentum and helicity of the fermion $l$ and neutrino, respectively,  
$(k,\sigma)$ is  the  four-momentum and helicity of the photon, and $(Q,\lambda)$ 
 the four-momentum and helicity of the $W$ boson.
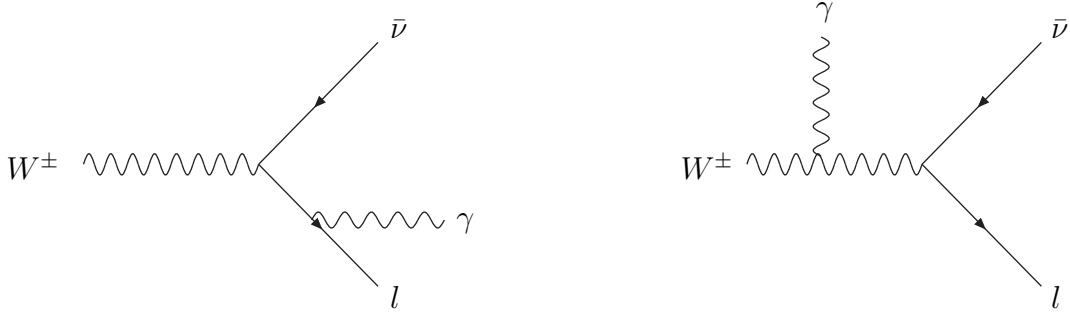
\begin{figure}[!ht] 
\begin{center}
 \begin{picture}(380,150)(-10,-30)
   \Photon(4,26)(70,26){4}{8}  \Text(-25,30)[lt]{$W^{\pm}$} 
   \ArrowLine(115,72)(70,26)   \Text(120,74)[lb]{$\bar{\nu}$}
   \ArrowLine(70,26)(115,-20)  \Text(120,-20)[lt]{${l}$}
   \Photon(90,5)(140,5){3}{5}  \Text(145.1,0)[lb]{$\gamma $}
   \Photon(254,26)(320,26){4}{8}  \Text(230,30)[lt]{$W^{\pm}$}
   \ArrowLine(365,72)(320,26)   \Text(370,74)[lb]{$\bar{\nu}$}
   \ArrowLine(320,26)(365,-20)  \Text(370,-20)[lt]{${l}$}
   \Photon(282,74)(282,29){3}{5} \Text(281,80)[lb]{$\gamma $}
 \end{picture}
\end{center}
\caption{ Feynman diagrams}
\label{feyndiag}  
\end{figure} 
\vspace{5cm}

\noindent In the notation of ref.~\cite{GPS}, the  corresponding parts of spin amplitudes take the  form:
\begin{eqnarray}
  \textit{M}^{\,\sigma}_{\lambda,\lambda_{\nu},\lambda_{l}}(k,Q,p_{\nu},p_{l}) &=& 
  \left[\frac{Q_{l}}{2\,k\cdot p_{l}}b_{\sigma}(k,p_{l}) -\frac{Q_{W}}{2\,k\cdot Q}\left(b_{\sigma}(k,p_{l})+b_{\sigma}(k,p_{\nu})\right) \right]{\textit{\large B}^{\lambda}_{\lambda_{l},\lambda_{\nu}}(p_{l},Q,p_{\nu})} \nonumber \\
 & & +\frac{Q_{l}}{2\,k\cdot p_{l}}\sum_{\rho=\pm}U^{\sigma}_{\lambda_{l},\rho}(p_{l},m_{l},k,0,k,0) {\textit{\large B}^{\lambda}_{\rho,-\lambda_{\nu}}(k,Q,p_{\nu})} \nonumber\\
 & &-\frac{Q_{W}}{2\,k\cdot Q}\sum_{\rho=\pm}\left({\textit{\large B}^{\lambda}_{\lambda_{l},-\rho}(p_{l},Q,k)} {U^{\sigma}_{-\rho,-\lambda_{\nu}}(k,0,k,0,p_{\nu},0)} \right. \\ \nonumber
 & & \left.\phantom{xxxxxxxxxx} + U^{\sigma}_{\lambda_{l},\rho}(p_{l},m_{l},k,0,k,0) {\textit{\large B}^{\lambda}_{\rho,-\lambda_{\nu}}(k,Q,p_{\nu})}\right), \nonumber
\end{eqnarray}
where we have introduced the following notation\,:
\begin{eqnarray}
 \textit{\large B}^{\lambda}_{\lambda_{1},\lambda_{2}}(p_{1},Q,p_{2})  
                     &\equiv& \frac{g}{2\sqrt{2}}\bar{u}(p_{1},\lambda_{1})\,\widehat{\epsilon}^{\lambda}_{W}(Q)(1+\gamma_{5})\,v(p_{2},\lambda_{2})\,,                                                    \nonumber \\
 U^{\sigma}_{\lambda_{1},\lambda_{2}}(p_{1},m_{1},k,0,p_{2},m_{2})
                     &\equiv& \bar{u}(p_{1},\lambda_{1})\,\widehat{\epsilon}^{\sigma}_{\gamma}(k)\,u(p_{2},\lambda_{2})\,,
                                                                                          \\
 \delta_{\lambda_{1}\lambda_{2}}b_{\sigma}(k,p) 
                     &\equiv& U^{\sigma}_{\lambda_{1},\lambda_{2}}(p,m,k,0,p,m) \nonumber \,,
\end{eqnarray}
 $Q_{l}$ and $Q_{W}$ are the electric charges of the fermion $l$ and the $W$ boson, respectively, 
in units of the positron charge, ${\epsilon}^{\sigma}_{\gamma}(k)$ and ${\epsilon}^{\lambda}_{W}(Q)$
denote respectively the polarization vectors of the photon and the $W$ boson. 

\begin{figure}[!ht]  
\setlength{\unitlength}{0.1mm}  
\begin{picture}(1600,600)  
\put( 375,750){\makebox(0,0)[b]{\large }}  
\put(1225,750){\makebox(0,0)[b]{\large }}  
\put( -60, 00){\makebox(0,0)[lb]{\epsfig{file=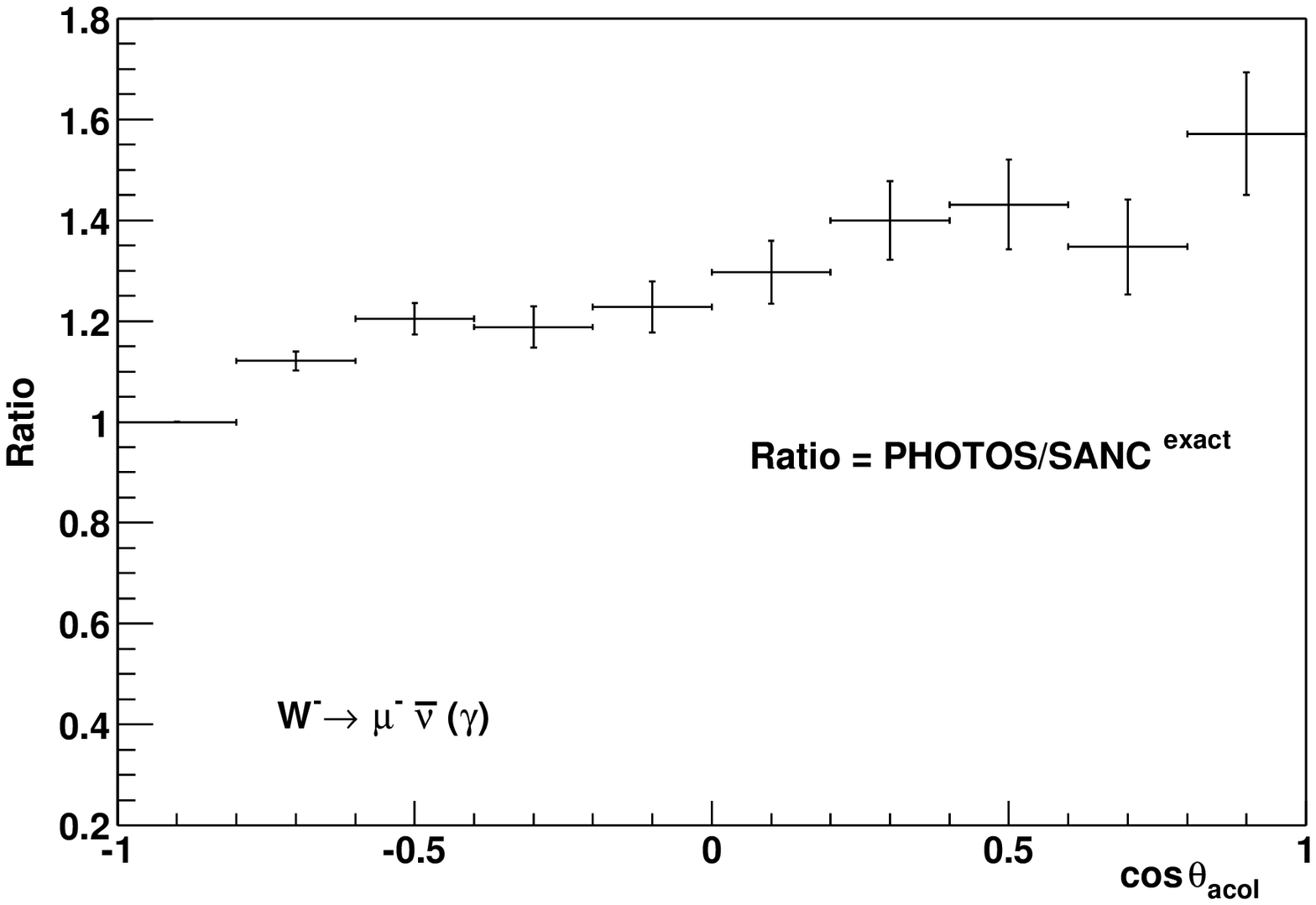,width=80mm,height=65mm}}}  
\put(860, 00){\makebox(0,0)[lb]{\epsfig{file=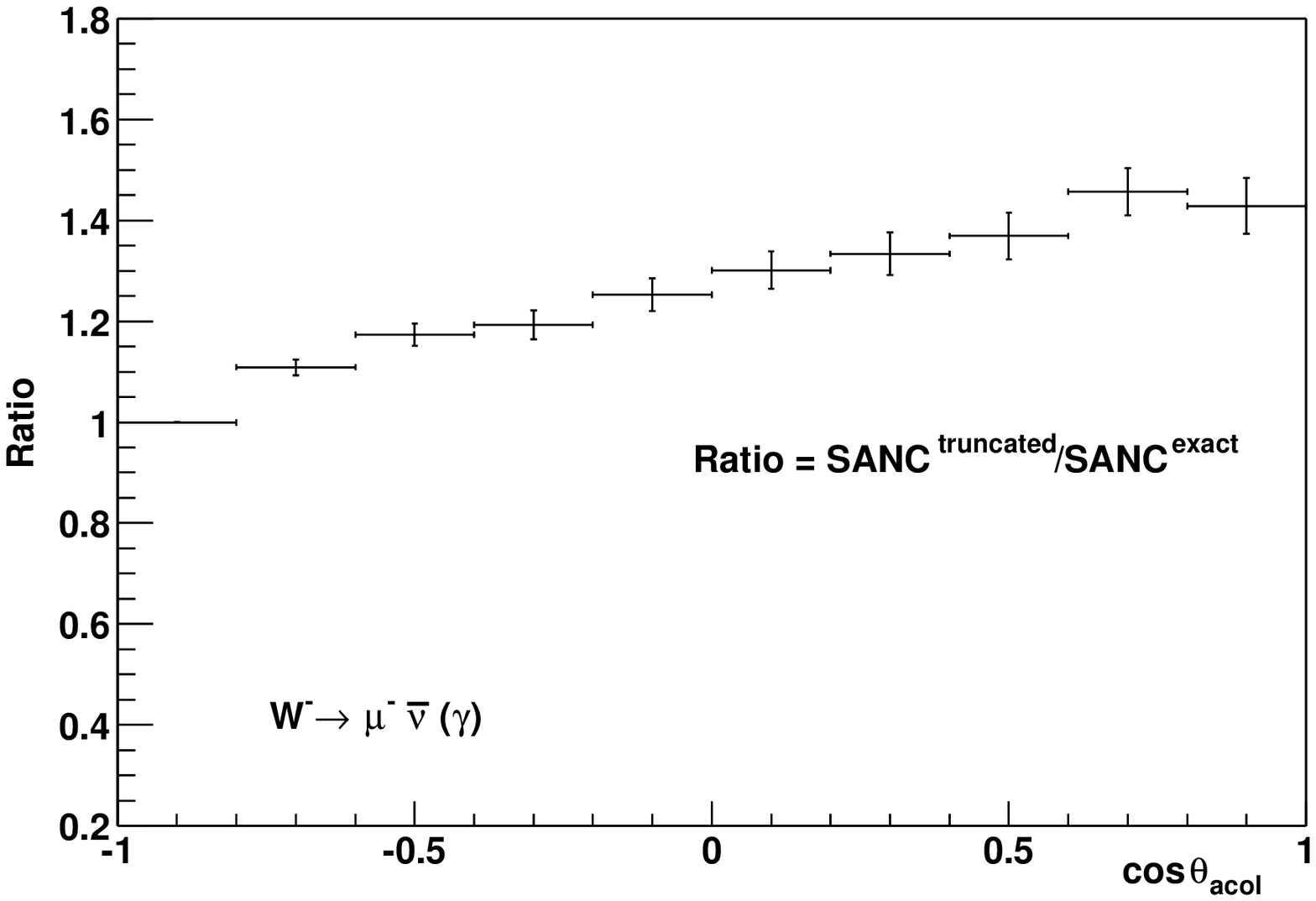,width=80mm,height=65mm}}}  
\end{picture}  
\caption  
{\it Comparisons (ratios) of the {\tt PHOTOS}, truncated \CCol \, and  complete \CCol \, predictions 
for the $W$ decay. Ratio of the $\mu^-\bar{\nu}$ acollinearity distribution from {\tt PHOTOS} 
and  complete \CCol\ (left-hand side) 
and ratio of the $\mu^-\bar{\nu}$ acollinearity distribution from  truncated \CCol
\, and  complete \CCol\ (right-hand side)  are given.
The dominant contribution is of non-leading  nature, the observable {\bf D}.}  
\label{bigdif}  
\end{figure}

This amplitude can be divided into three gauge-invariant parts:
\begin{itemize}
 \item Infrared-divergent part:\\
    \begin{eqnarray}
       \textit{M}^{\,\sigma}_{\lambda,\lambda_{\nu},\lambda_{l}}(k,Q,p_{\nu},p_{l})^{(a)} &=& 
        \left[\frac{Q_{l}}{2\,k\cdot p_{l}}b_{\sigma}(k,p_{l}) \right. \\ \nonumber
        && \left.\phantom{xx} -\frac{Q_{W}}{2\,k\cdot Q}\left(b_{\sigma}(k,p_{l})+b_{\sigma}(k,p_{\nu})\right) \right]{\textit{\large B}^{\lambda}_{\lambda_{l},\lambda_{\nu}}(p_{l},Q,p_{\nu})} \nonumber
    \end{eqnarray}      
 \item Collinear-divergent part:\\
    \begin{eqnarray}
     \textit{M}^{\,\sigma}_{\lambda,\lambda_{\nu},\lambda_{l}}(k,Q,p_{\nu},p_{l})^{(b)} &=& 
      \frac{Q_{l}}{2\,k\cdot p_{l}}\sum_{\rho=\pm}U^{\sigma}_{\lambda_{l},\rho}(p_{l},m_{l},k,0,k,0) {\textit{\large B}^{\lambda}_{\rho,-\lambda_{\nu}}(k,Q,p_{\nu})} 
    \end{eqnarray}      
 \item Finite part : \\
    \begin{eqnarray}
     \lefteqn{\textit{M}^{\,\sigma}_{\lambda,\lambda_{\nu},\lambda_{l}}(k,Q,p_{\nu},p_{l})^{(c)} =
     -\frac{Q_{W}}{2\,k\cdot Q}\sum_{\rho=\pm}\left({\textit{\large B}^{\lambda}_{\lambda_{l},-\rho}(p_{l},Q,k)} {U^{\sigma}_{-\rho,-\lambda_{\nu}}(k,0,k,0,p_{\nu},0)}\right.}\\
   &&\left. \phantom{xxxxxxxxxxxxxxxxxxxxxxxxx}+ U^{\sigma}_{\lambda_{l},\rho}(p_{l},m_{l},k,0,k,0) {\textit{\large B}^{\lambda}_{\rho,-\lambda_{\nu}}(k,Q,p_{\nu})}\right). \nonumber
    \end{eqnarray}      
\end{itemize}
Once the last (finite) term  is dropped  from the whole amplitude, 
event distributions produced with {\tt SANC} and {\tt PHOTOS} become almost identical; this is shown in fig. 
\ref{bigdif}, where comparisons of the {\tt PHOTOS} (truncated {\tt SANC}) with complete matrix elements 
are visible on the left-hand (right-hand) part of the figure.
This is very encouraging and can  be used as a starting point for improving {\tt PHOTOS},
with the help of some kind  of  correction  weight. We have indeed 
found that analysing the difference between the complete and truncated matrix elements of 
{\tt SANC} such a correction weight $\delta$ can be found. One can replace 
the exact matrix element of {\tt SANC} with: 

\begin{eqnarray}
  \sum_{\sigma\lambda\lambda_{\nu}\lambda_{l}}|\textit{M}^{\,\sigma}_{\lambda,\lambda_{\nu},\lambda_{l}}(k,Q,p_{\nu},p_{l})^{(a)}+ \textit{M}^{\,\sigma}_{\lambda,\lambda_{\nu},\lambda_{l}}(k,Q,p_{\nu},p_{l})^{(b)}|^2 \left( 1+\delta\right),  
\end{eqnarray}

\begin{figure}[!ht]  
\setlength{\unitlength}{0.1mm}  
\begin{picture}(1600,600)  
\put( 375,750){\makebox(0,0)[b]{\large }}  
\put(1225,750){\makebox(0,0)[b]{\large }}  
\put( -60, 00){\makebox(0,0)[lb]{\epsfig{file=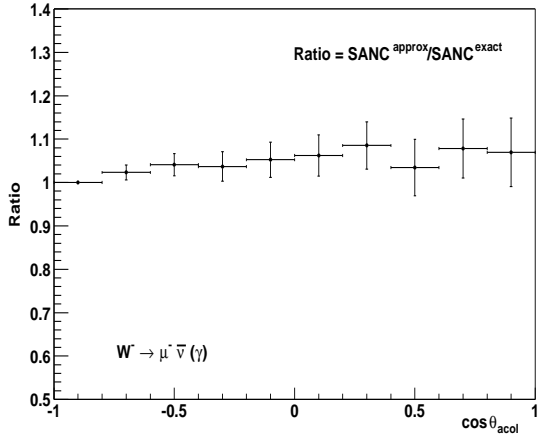,width=80mm,height=65mm}}}  
\end{picture}  
\caption  
{\it Comparison (ratio) of the predictions 
for the $W$ decay from \CCol\  truncated and corrected with the function $\delta$ and from the 
complete \CCol{}. Ratio of the $\mu^-\bar{\nu}$ acollinearity distribution is given.
The dominant contribution is of non-leading  nature, the observable {\bf D}.}  
\label{deltafig}  
\end{figure}

where 
\begin{eqnarray}
   \delta &=& \frac{\sum_{\sigma\lambda\lambda_{\nu}\lambda_{l}}|\textit{M}^{\,\sigma}_{\lambda,\lambda_{\nu},\lambda_{l}}(k,Q,p_{\nu},p_{l})|^2 }{\sum_{\sigma\lambda\lambda_{\nu}\lambda_{l}}|\textit{M}^{\,\sigma}_{\lambda,\lambda_{\nu},\lambda_{l}}(k,Q,p_{\nu},p_{l})^{(a)} + \textit{M}^{\,\sigma}_{\lambda,\lambda_{\nu},\lambda_{l}}(k,Q,p_{\nu},p_{l})^{(b)}|^2} -1  \nonumber\\
          &=& -8 k_{0}(1-\beta_{l}\cos\theta)
                      \left(\frac{m_{l}^{2}E_{l}}{M_{W}^{4}(1-m_{l}^{2}/M_{W}^{2})(4-m_{l}^{2}/M_{W}^{2})}\right.\nonumber                                              \\
          & &               \left.\phantom{xxxxxxxxxxxxxxxxx}+\frac{2E_{l}k_{0}}{M_{W}^{3}(1-m_{l}^{2}/M_{W}^{2})(4-m_{l}^{2}/M_{W}^{2})} \right) 
\label{delta}
\end{eqnarray}
and $k_{0}$ denotes the photon energy, while $m_{l}$ and $E_{l}$ denote respectively 
the mass and the energy of the fermion $l$. Finally
$M_{W}$ denotes the mass of the $W$ boson and 
$\beta_{l}=\sqrt{1-m_{l}^{2}/E_{l}^{2}}$, $\theta=\angle (\vec{p_{l}},\vec{k})$ (all 
kinematical variables defined
in the $W$ rest frame). 
The differences between the  results obtained from such approximated matrix elements and those 
from the complete matrix elements  are rather small (see fig.~\ref{deltafig}).

\section{ Numerical results from the improved {\tt PHOTOS}}

Once our formula (\ref{delta}) was expressed in a simple form, we could rather easily introduce it 
into {\tt PHOTOS} as an additional correcting weight, to be activated in the case of leptonic $W$ decays only.

\begin{figure}[!ht]  
\setlength{\unitlength}{0.1mm}  
\begin{picture}(1600,830)  
\put( 375,750){\makebox(0,0)[b]{\large }}  
\put(1225,750){\makebox(0,0)[b]{\large }}  
\put( -60, 00){\makebox(0,0)[lb]{\epsfig{file=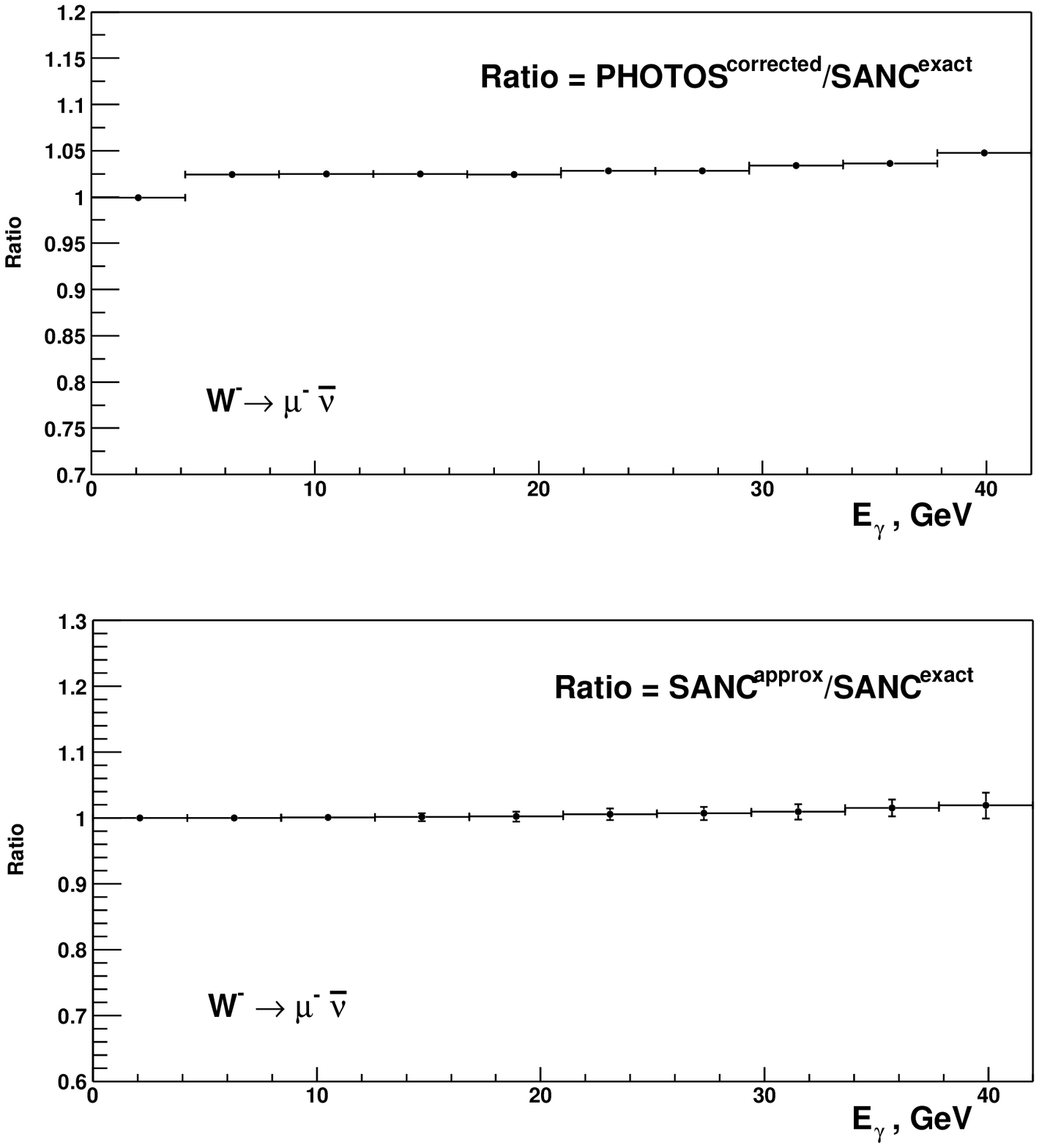,width=80mm,height=85mm}}}  
\put(860, 00){\makebox(0,0)[lb]{\epsfig{file=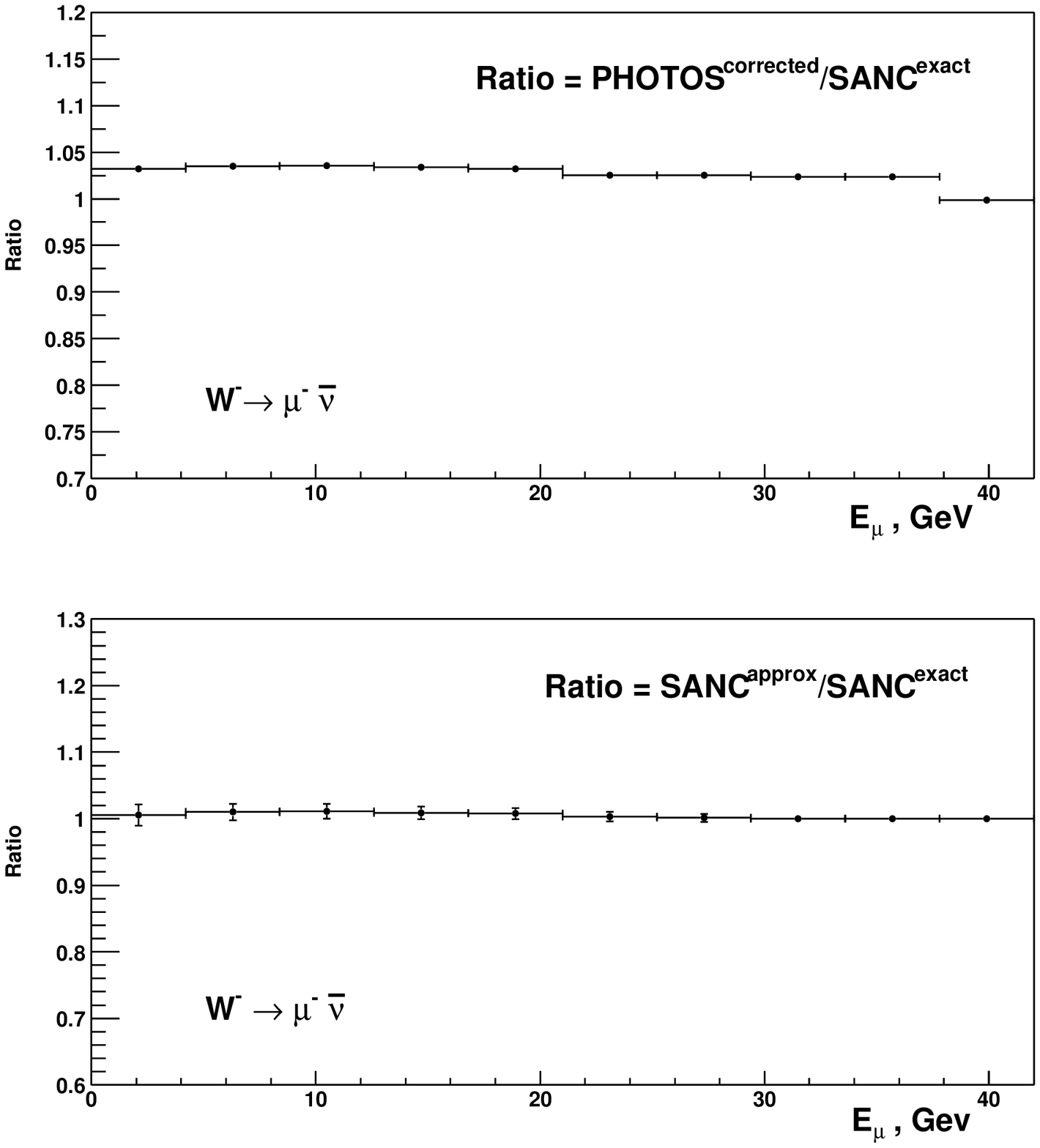,width=80mm,height=85mm}}}  
\end{picture}  
\caption  
{\it Comparisons (ratios) of the  complete  \CCol\ and corrected {\tt PHOTOS}  predictions 
for the $W$ decay. Observables {\bf A} and {\bf B}: ratios of the photon energy (left-hand side) 
and muon energy (right-hand side)  distributions from the two programs.
The dominant contribution is of  leading-log (collinear) nature.  
 In the lower part of the
plots similar comparisons for the complete  \CCol\ and truncated--corrected with $\delta$ \CCol\ 
predictions are given. } 
\label{figAB-W}  
\end{figure}

\begin{figure}[!ht]  
\setlength{\unitlength}{0.1mm}  
\begin{picture}(1600,830)  
\put( 375,750){\makebox(0,0)[b]{\large }}  
\put(1225,750){\makebox(0,0)[b]{\large }}  
\put( -60, 00){\makebox(0,0)[lb]{\epsfig{file=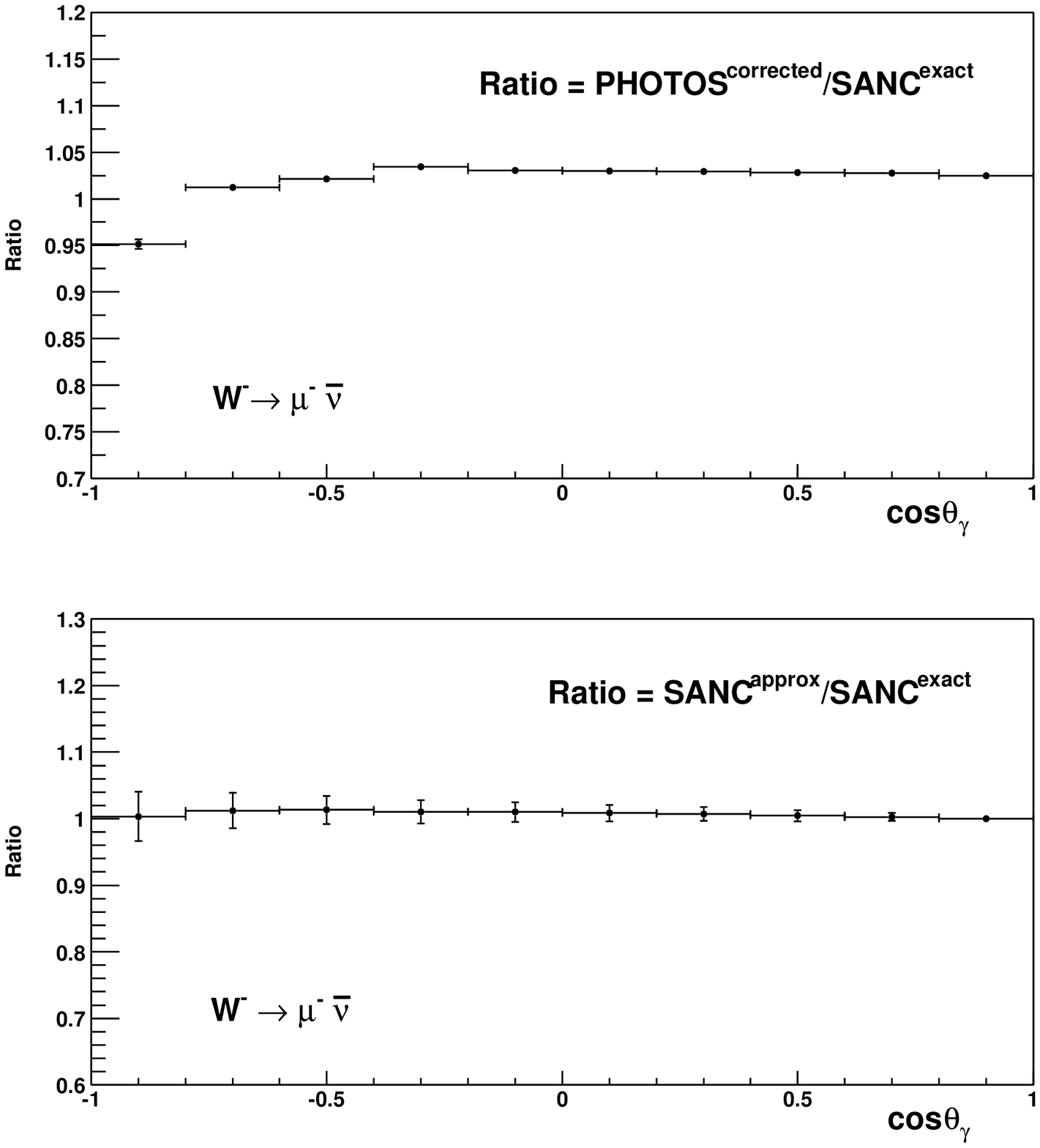,width=80mm,height=85mm}}}  
\put(860, 00){\makebox(0,0)[lb]{\epsfig{file=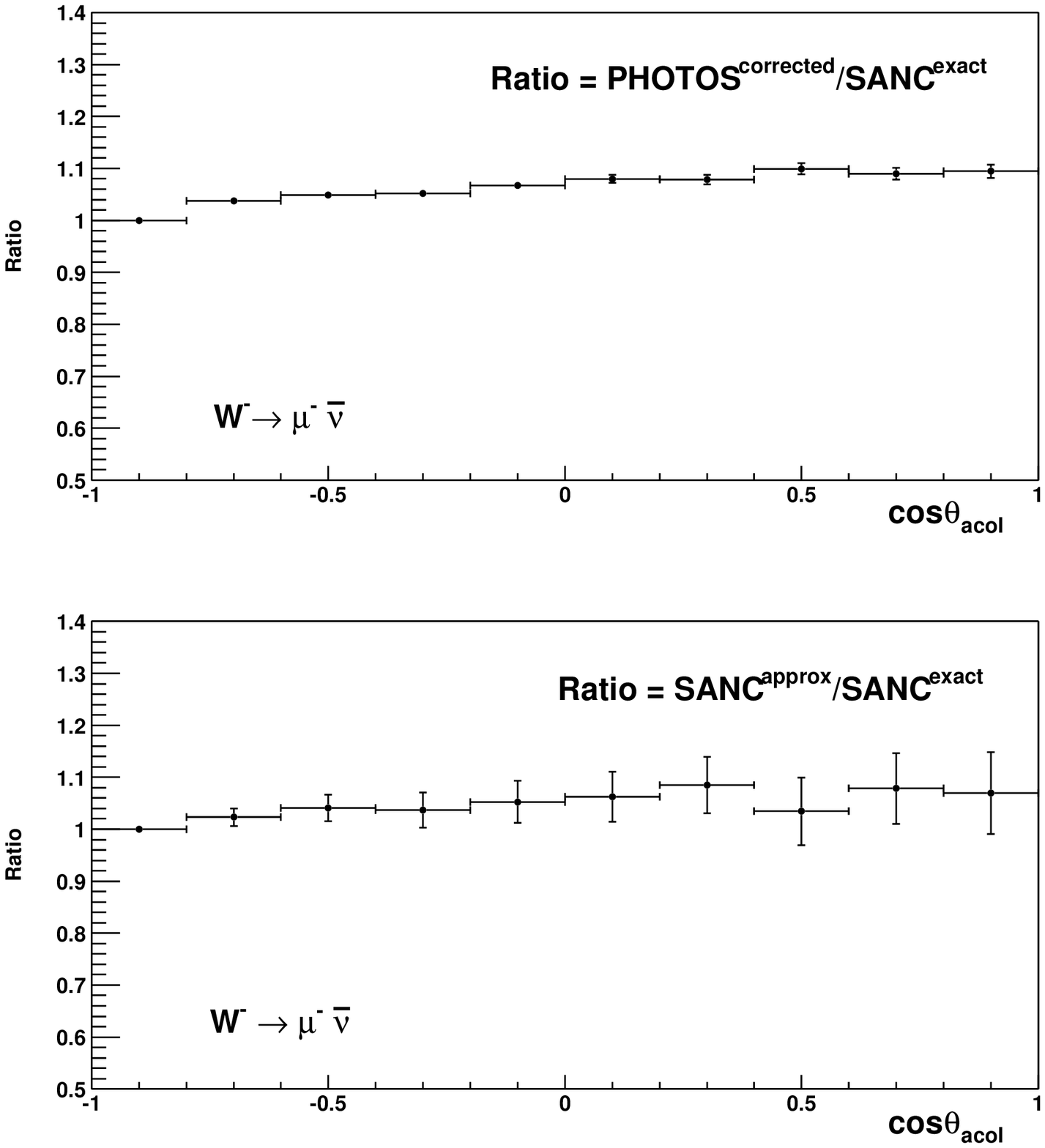,width=80mm,height=85mm}}}  
\end{picture}  
\caption  
{\it Comparisons (ratios) of the  complete  \CCol\ and corrected {\tt PHOTOS}  predictions 
for the $W$ decay. Observables {\bf C} and {\bf D}: ratios of the photon angle with respect 
to $\mu^-$ (left-hand side) 
and $\mu^-\bar{\nu}$ acollinearity (right-hand side)  distributions from the two programs.
The dominant contribution is of infrared non-leading-log  nature for the left-hand side plot,
and non-infrared non-leading-log nature for the right-hand side one. In the lower part of the
plots similar comparisons for the complete  \CCol\ and truncated--corrected with $\delta$ \CCol\ 
predictions are given. } 

\label{figCD-W}  
\end{figure}

As we can see from figs. \ref{figAB-W}, and  \ref{figCD-W}  
level of the agreement between the new version of {\tt PHOTOS} and the complete-matrix-elements
 calculation 
of   \CCol\ is for all distributions of the types {\bf A, B, C} and
{\bf D}  at the level 
of better than 10\%{}. This is true, even in the case of the observable {\bf D}, 
where the differences, without the correcting weights,
were up to 40\%. 

Also, as one can see  in fig.   \ref{figCD-W},
comparisons between complete and 
corrected--truncated versions of  \CCol\  give similar patterns of residual differences, 
as in case of {\tt PHOTOS} comparisons with the complete matrix elements \CCol{}. 
This indicates that the approximation used in evaluating formula (\ref{delta}) is
at the origin of at least some of the differences between corrected  {\tt PHOTOS} 
and full-matrix-element 
calculation, rather than e.g. some
technical problems related to its Monte Carlo algorithm.

With the correcting weight (\ref{delta}) included, the comparisons of {\tt PHOTOS}
with the ``matrix-elements'' type calculations improved significantly.
The differences were not only largely removed, but also explained as being due to part of 
the $W$--$W$--$\gamma$ interaction explicitly missing in older versions of {\tt PHOTOS}.
This indicates that, in future, algorithms of this type can be improved even further,
e.g. using exponentiation techniques. On the other hand, differences between lower and upper 
plots
of fig. \ref{figAB-W}  point toward the limits of the method.

At LEP2, the production and decay of $W$  pairs is now being combined
for all four LEP2 experiments, and
uncertainties due to bremsstrahlung in $W$ decay are important.
{\tt PHOTOS} is part of the  main programs (see e.g. YFSWW3 \cite{Jadach:2001mp})
used in the LEP2 analysis of the  $W$-pair data.
The present paper provides a means of  estimating the size of the uncertainties from
{\tt PHOTOS}. The correction weight can be switched on and off for that purpose.
 We hope that, for LEP2
experiments, the updated version of {\tt PHOTOS} will turn out to be sufficient. However,  
for  higher
precision, more sophisticated solutions, e.g. as those described in ref.~\cite{Placzek:2003zg},  
will have to be used.
Finally, let us mention that
the $W$ channel  is of importance for some studies 
of the LHC Higgs-discovery potential as well \cite{Richter-Was:1993ta,ATLASTDR}.

\section{Summary}

We have successfully tested \CCol\ versus {\tt PHOTOS}
in the case of the $W$ decay.
We have first checked that once non-leading contributions of \CCol\ 
related to the $W$ charge are dropped,  \CCol\ tends to  agree
 with the results 
provided by the old version of {\tt PHOTOS} Monte Carlo. 
These comparisons convinced  us that the deficiencies of {\tt PHOTOS} are
indeed due to missing non-leading terms. The missing terms were then
studied analytically within \CCol\ and an approximating correcting weight, 
for leptonic $W$ decays, was
found. Once the weight was installed to the new version of  {\tt PHOTOS},
agreement with respect to complete results of  \CCol\ was found. With new
correcting weight, the
 {\tt PHOTOS} predictions are within 5\%    (instead of 7\%  without
correcting weight)  for the 
end parts of the spectra affected by the leading-log corrections, but 
within 5\% (instead of 20\%) for the angular part of the distributions, where the 
infrared-induced logarithms dominate.
The agreement is now also at the 10\% level  in the phase-space  regions 
where only non-leading corrections contribute to the matrix elements. 
This is a significant improvement with respect to old version
of {\tt PHOTOS}, without correcting weight, where differences were up to 40\%{}. 

\section*{Acknowledgements}
The authors would like to thank D. Bardin, F. Cossutti, S. Jadach and B.F.L. Ward
for useful discussions.



\begin{thebibliography}{99}
 
\bibitem{Bardin:2002am}
D.~Bardin {\em et~al.}, 
``Project {SANC} (former {CalcPHEP}): Support of analytic and numeric calculations for experiments at colliders'',
CERN-TH/2002-245, {\tt arXiv:hep-ph/0209297}, to be published in ICHEP2002 Proceedings; \\
D.~Bardin {\em et~al.}, 
``Project {\tt CalcPHEP}: Calculus for precision high energy physics'',
{\tt arXiv:hep-ph/0202004}, CAAP-2001 Proceedings, Dubna 2001;\\
D.~Bardin, P.~Christova, L.~Kalinovskaya, ``{\tt SANC} Status Report",
{\it Nucl. Phys. B (Proc. Suppl.)} {\bf 116} (2003) 48--52.

\bibitem{Gizo}
 A.~Andonov, S.~Jadach, G.~Nanava and Z.~W\c{a}s, ``Comparison of {\tt SANC} with {\tt KORLAZ} and {\tt PHOTOS}'', CERN-TH-2002-315, {\tt arXiv:hep-ph/0212209},
 to be published {\em in Acta Physica Polonica}.

\bibitem{Kobel:2000aw}
M.~Kobel {\em et~al.}, ``Two Fermion Working Group Collaboration'', {\tt hep-ph/0007180.}

\bibitem{Barberio:1994qi}
E.~Barberio and Z.~W\c{a}s,
{\em Comput. Phys. Commun.} {\bf 79} (1994) 291.

\bibitem{Barberio:1991ms}
E.~Barberio, B.~van ~Eijk and Z.~W\c{a}s,
{\em Comput. Phys. Commun.} {\bf 66} (1991) 115.

\bibitem{GPS}
S.~Jadach, B.F.L.~Ward  and Z.~W\c{a}s,
{\em Phys. Lett.} {\bf B449} (1999) 97.


\bibitem{Richter-Was:1993ta}
E.~Richter-W\c{a}s,
Z.\ Phys.\ C {\bf 61} (1994) 323.

\bibitem{ATLASTDR}  ATLAS collaboration, 
`` ATLAS Detector and Physics Performance Technical Design Report'', 
CERN/LHCC/99-15, vol. 2, p. 674--735.
\bibitem{Jadach:2001mp}
S.~Jadach, W.~Placzek, M.~Skrzypek, B.~F.~Ward and Z.~Was,
{\em Comput.\ Phys.\ Commun.}\  {\bf 140} (2001)  475,
arXiv:hep-ph/0104049.
\bibitem{Placzek:2003zg}
W.~Placzek and S.~Jadach,
``Multiphoton radiation in leptonic W-boson decays,''
arXiv:hep-ph/0302065.
\end{thebibliography}

\end{document}